%% file: paper.tex
\documentclass[twocolumn,showpacs,aps,prl,superscriptaddress,letterpaper]{revtex4}

\usepackage{graphicx}
\usepackage{dcolumn}
\usepackage{amsmath}
\usepackage{epsfig}

\input pubboard/babarsym

\def\signal {\tau^- \rightarrow  3\pi^- 2\pi^+ 2\pi^0  \nu_\tau }
\def\fivepipizero {\tau^- \rightarrow  3\pi^- 2\pi^+ \pi^0 \nu_\tau }

\def\twopizero {\tau^- \rightarrow 2\pi^- \pi^+ 2\pi^0  \nu_\tau }
\def\threepizero {\tau^- \rightarrow  2\pi^- \pi^+ 3\pi^0  \nu_\tau }
\def\twoomegapi {\tau^- \rightarrow  2\omega \pi^- \nu_\tau }
\def\omegathreepi {\tau^- \rightarrow  \omega 2\pi^-  \pi^+ \nu_\tau }
\def\omegadec {\omega \rightarrow  \pi^-\pi^+\pi^0 }

\newcommand\beq{\begin{equation}}
\newcommand\eeq{\end{equation}}
\newcommand\ov{\overline}

\def\eetoqq {e^+e^-\rightarrow q\bar q}
\def\eetocc {e^+e^-\rightarrow c\bar c}
\def\eetouds {e^+e^-\rightarrow (u\bar u, d\bar d,  s\bar s)}

\newcommand{\BABARPubYear}    {06}
\newcommand{\BABARPubNumber}  {008}

\newcommand{\SLACPubNumber} {11804}

\newcommand{\lumi}    {232\invfb}

\def\figurebox#1#2#3{%
    \def\arg{#3}%
    \ifx\arg\empty
    {\hfill\vbox{\hsize#2\hrule\hbox to #2{\vrule\hfill\vbox to #1{\hsize#2\vfill}\vrule}\hrule}\hfill}%
    \else
    {\hfill\epsfbox{#3}\hfill}%
    \fi}

\long\def\inst#1{\par\nobreak\kern 4pt\nobreak
    {\it #1}\par\vskip 10pt plus 3pt minus 3pt}

\begin{document}
\bibliographystyle{prsty}

\preprint{\babar-PUB-\BABARPubYear/\BABARPubNumber} 
\preprint{SLAC-PUB-\SLACPubNumber} 

\begin{flushleft}
\babar-PUB-\BABARPubYear/\BABARPubNumber\\
SLAC-PUB-\SLACPubNumber\\
\end{flushleft}

\title{
{\large \bf Search for the Decay {\boldmath $\signal$}} 
}

\input pubboard/authors_feb2006.tex

\begin{abstract}
A search for the decay of the $\tau$ lepton to five charged and two neutral pions
is performed using data collected by the \babar\ detector at
the PEP-II asymmetric-energy $e^{+}e^{-}$ collider.
The analysis uses \lumi of data at center-of-mass energies on or near the
\FourS resonance.
We observe 10 events with an expected background of $6.5^{+2.0}_{-1.4}$ events.
In the absence of a signal, we set the limit on the branching ratio
\BR$(\signal) < 3.4\times10^{-6}$ at the 90\,\% confidence level. 
This is a significant improvement over the previously established limit.
In addition, we search for the decay mode $\twoomegapi$.
We observe 1 event with an expected background of 0.4$^{+1.0}_{-0.4}$ events
and calculate the upper limit \BR$(\twoomegapi) < 5.4\times10^{-7}$ at the 90\,\% confidence level. 
This is the first upper limit for this mode.
\end{abstract}


\maketitle
Hadronic decays of $\tau$ leptons provide an excellent laboratory for the study
of the strong interaction. Decays of the $\tau$ with one or three charged particles
in the final state have been well studied in the past~\cite{pdg04}.
Higher multiplicity decays, however, have considerably lower branching ratios~\cite{pdg04},
and high luminosity experiments are needed to study their dynamics and search for new modes.
The \babar\ experiment has recorded a large sample of $e^+e^- \rightarrow \tau^+\tau^-$ 
events suitable for detailed searches for high multiplicity $\tau$ decays.

The $\signal$ mode~\cite{footnote} is of particular interest. It has not been observed yet,
and only an upper limit \BR$(\signal) < 1.1\times10^{-4}$ at the 90\,\% confidence level (CL) has been set
by the CLEO collaboration~\cite{CLEO}. The reason for the suppression
of seven-pion $\tau$ decays is the limited phase space of this decay~\cite{Nussinov, 7prong}. 
For the same reason, if this decay is observed with sufficient statistics, it may lead to a more stringent limit
on the $\tau$ neutrino mass.

Since $\tau$ decays to five charged pions and a $\pi^{0}$ meson 
most likely involve resonances (e.g., $\omega$ or $\eta$)~\cite{5pipi0}, it is expected that the $\signal$
decay may also proceed through resonances.
According to calculations based on isospin symmetry~\cite{Sobie}, 
the decay $\twoomegapi$ is expected to be the dominant mode.

This analysis is based on data recorded with
the \babar\ detector at the \pep2\ asymmetric-energy \epem storage 
ring operated at the Stanford Linear Accelerator Center.
The data sample consists of \lumi recorded at center-of-mass (CM) energies
of $10.58 \gev$ and $10.54 \gev$.
With an expected cross section for $\tau$ pairs of $\sigma_{\tau\tau} = (0.89\pm0.02)$\nb~\cite{kk},
the number of produced $\tau$-pair events is $N_{\tau\tau} = (206.5\pm4.7)\times 10^6$.

The \babar\ detector is described in detail in Ref.~\cite{detector},
and only a brief description is given here.
Charged-particle momenta are measured with a 5-layer
double-sided silicon vertex tracker (SVT) and a 40-layer drift chamber (DCH) 
inside a solenoidal magnet with a 1.5 T magnetic field.
A calorimeter (EMC) consisting of 6580 CsI(Tl) 
crystals is used to measure the energy of electrons, positrons, and photons.
A ring-imaging Cherenkov detector is used to identify
charged hadrons, in combination with ionization energy loss
measurements in the SVT and the DCH.
Muons are identified by an instrumented magnetic-flux return (IFR).

Monte Carlo (MC) simulations are used to estimate the $\signal$ signal efficiency
and background contamination from other $\tau$ decay modes.
The production of $\tau$ pairs is simulated with the {\mbox{\tt KK}\xspace} generator~\cite{KK2f},
and non-signal $\tau$ lepton decays are modeled with {\mbox{\tt TAUOLA}\xspace}~\cite{TAUOLA} according
to measured rates~\cite{pdg04}. The background processes $\eetoqq$ ($q = u,d,s,c,b$) are simulated
using the {\mbox{\tt JetSet}\xspace} package~\cite{JetSet}.
Signal events are generated using phase space with a $V-A$ interaction. 
We find no significant variation in efficiency within the phase space. 
The simulation of the \babar\ detector is based on \geant~4~\cite{geant4}.

The principal backgrounds to our signal come from $\eetoqq$ processes and 
multi-pion $\tau$ decay modes involving at least one $\pi^0$, namely
$\fivepipizero$, $\twopizero$ and $\threepizero$ modes. 
The $\fivepipizero$ contribution comes from reconstructing an additional (fake) $\pi^0$,
while the three-prong modes contribute through the $\pi^0$ decay to a photon pair 
and subsequent photon conversions in detector material.

The event selection criteria were developed to suppress the background while
maintaining high signal efficiency.
Events with six charged particle tracks and a net charge of zero are first selected.
To ensure well-reconstructed tracks, each track is required to have a minimum transverse momentum of $100\,\mevc$,
a distance of closest approach to the interaction point in the
plane transverse to the beam axis (DOCA$\rm_{XY}$) less than
1.5\,cm, and a distance of closest approach along the beam direction
less than 10\,cm.
Four or more tracks are required to have hits in at least 12 DCH layers.
Photons are reconstructed from clusters in the EMC and are required to have a minimum energy 
of $50\,\mev$, energy deposited in at least three crystals, and a lateral energy profile consistent with that of a photon.
In addition, to suppress background from backscattering in the EMC, the angle between the position of 
a cluster and the impact point
of the nearest charged track at the EMC surface, as seen from the interaction point, 
is required to be more than 0.08 radians.

The $\pi^0$ mesons are reconstructed from two photon candidates passing the photon selection criteria
described above. 
We first search for $\pi^0$ candidates with energy $E_{\pi^0}>450\,\mev$ and mass $113<M_{\gamma\gamma}<155\,\mevcc$.
If two or more $\pi^0$ candidates share a photon, only the
one with the smallest $|M_{\gamma\gamma} - M_{\pi^0}^{\rm PDG}|$, where $M_{\pi^0}^{\rm PDG}$ value is taken
from~\cite{pdg04}, is retained. 
Next, we repeat the procedure for $\pi^0$ candidates
with energy $300<E_{\pi^0}<450\,\mev$ and mass $120<M_{\gamma\gamma}<148\,\mevcc$.

The $\tau$ pair is produced approximately back-to-back in the \epem CM
frame. This allows the event to be divided into two hemispheres by a plane perpendicular to the thrust
axis, where the thrust is calculated from all charged particles and photons in
the event~\cite{JetSet}. The event thrust magnitude is required to be larger than 0.9.
This requirement rejects more than 90\,\% of the $q\ov{q}$ background and
the $e^+e^- \rightarrow B\ov{B}$ background is suppressed to a negligible level.
Events are required to have one track in one hemisphere (the tag side)
and five tracks in the other hemisphere (the signal side).
To further suppress the background from $\eetoqq$ events,
we demand a well-identified electron or muon on the tag side with at most one additional
photon with energy $E_{\gamma}<500\,\mev$. 
The combined mass of all charged particles and photons in each hemisphere is required to be less than $3\,\gevcc$. 
Finally, only events with exactly two $\pi^0$ candidates on the signal side are kept for further study.
The efficiency of the two $\pi^0$ selection in the signal MC is 13.0\,\%.

The visible energy, defined as the sum of the CM energy of
the charged tracks and the reconstructed $\pi^0$ mesons, is required to be less
than the CM beam energy $E_{\rm beam}=5.29\,\gev$ in each hemisphere of the event.
The residual energy $E_{\rm res}$, defined as the neutral energy on the signal side not associated with the
reconstructed $\tau$ decay products, is required to be less than $300\,\mev$, 
reducing the background from $\eetoqq$ and $\threepizero$ events.

\begin{figure}[bht]
\begin{center}
\epsfig{file=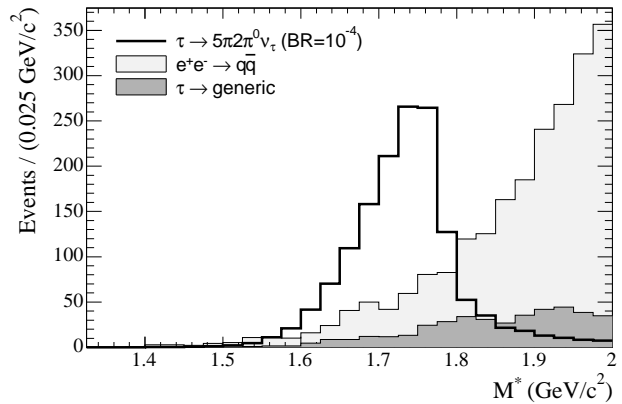,width=0.47\textwidth}
\caption{\small{Pseudo-mass distribution below $2\,\gevcc$ for signal and background 
MC samples. Signal is plotted assuming \BR$(\signal) = 10^{-4}$. 
For illustrative purposes the 1-prong tagging requirements are not imposed here.
\label{fig:pseudomass}}} 

\end{center}
\end{figure}

To reconstruct the signal event, an approximation of the $\tau$ invariant mass is used:
\begin{equation}
M^{*2} = 2\,(E_{\rm beam} - E_{7h})(E_{7h} - P_{7h}) + M_{7h}^{2},
\end{equation}
where the $\tau$ neutrino is assumed to be massless and travel along the direction of the combined 
momentum vector $P_{7h}$ of the seven hadrons and its energy is taken to be the difference 
between $E_{\rm beam}$ and the combined energy $E_{7h}$ of the hadrons in the CM system.
The variable $M^{*}$ is called the $\tau$ pseudo mass~\cite{pseudomass},
and its distributions for signal and background MC events are shown in Figure~\ref{fig:pseudomass}.
The advantage of $M^*$ over the invariant mass $M_{7h}$ is a considerably better
separation of the signal from the hadronic $q\ov{q}$ background.

We apply particle identification on the signal side,
demanding four out of five tracks to be identified as pions with high probability, and apply
looser identification criteria to the fifth track. This requirement significantly
reduces the background from $\tau$ events with photon conversions
and $\eetoqq$ events containing kaons.

We further suppress photon conversions by requiring
the invariant mass of each pair of oppositely charged tracks
to be larger than $5\,\mevcc$.
In addition, we apply cuts on the sums of the two lowest transverse momenta and two largest DOCA$\rm_{XY}$
of the tracks on the signal side: $p_t^{\rm lowest1} + p_t^{\rm lowest2}>0.4\,\gevc$
and DOCA$\rm_{XY}^{\rm largest1}$ + DOCA$\rm_{XY}^{\rm largest2}<0.4\,\cm$.

The final event count is performed in the signal region $1.3<M^{*}<1.8\,\gevcc$.
According to MC studies, the signal efficiency after all cuts is $(0.66\pm0.05)\,\%$.
The error is a combination of systematic and statistical uncertainties.
The systematic uncertainty on the signal efficiency includes contributions
from the reconstruction of charged tracks and photons (4.3\,\%), the reconstruction of two $\pi^0$ mesons (6.6\,\%), 
and the uncertainty associated with the particle identification on the signal and tag sides (1.7\,\%). 
A statistical uncertainty (1.8\,\%) due to limited MC
samples is added in quadrature to the systematic uncertainty.

\begin{figure}[t]
\begin{center}
\begin{minipage}{0.235\textwidth}
\epsfig{file=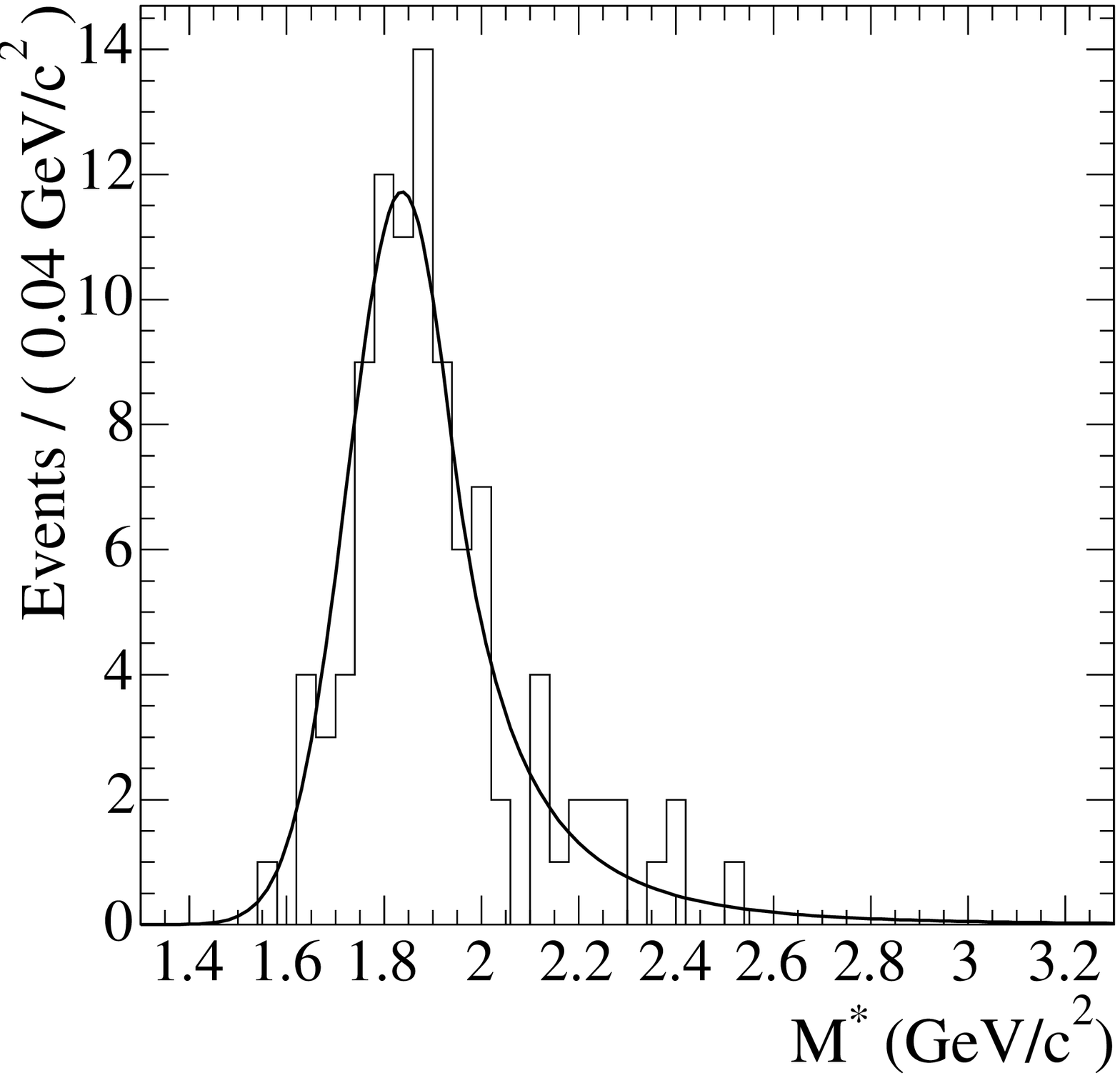,width=1.0\textwidth}
\end{minipage}
\begin{minipage}{0.235\textwidth}
\epsfig{file=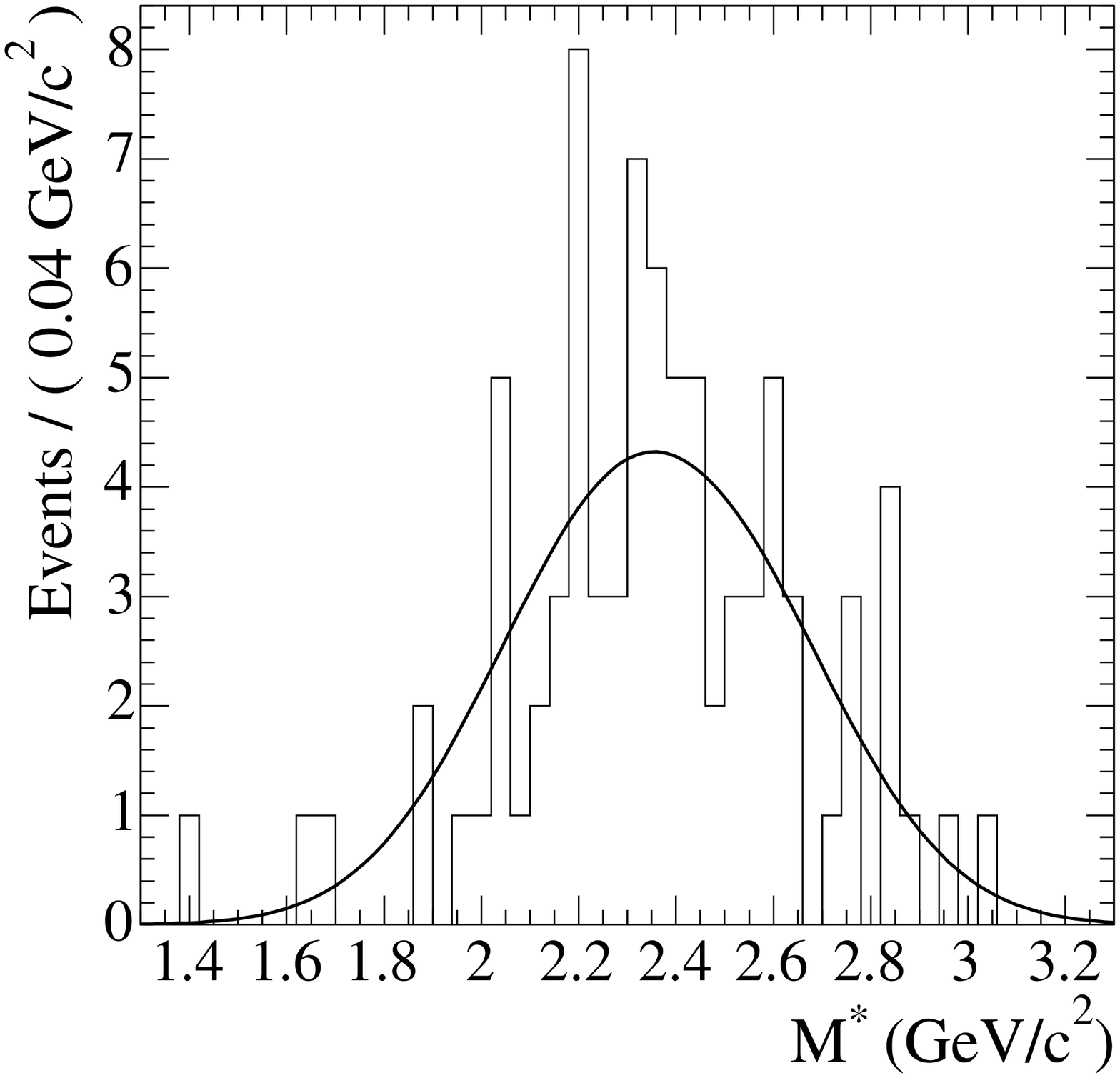,width=1.0\textwidth}
\end{minipage}
\caption{\small{Monte Carlo simulated pseudo-mass distributions of the $\fivepipizero$ 
background with a ``Crystal Ball'' shape PDF superimposed (left)
and $\eetoqq$ background fitted with the sum of two Gaussians (right).
The distributions are not normalized to the data luminosity. \label{fig:fitbkg}}} 
\end{center}
\end{figure}

The simulation of $\tau$-pair events yields a reliable
estimate of their expected background contribution, verified by 
modifying the event selection criteria to suppress the $q\ov{q}$ background and allow for more $\tau$ events.
The largest background is predicted to come from $\fivepipizero$ decays. 
For a detailed study, we use an MC sample of $\fivepipizero$ events corresponding to 
1900\,$\invfb$ of data. The pseudo-mass
spectrum of the events passing the selection criteria is fitted with a ``Crystal Ball''
probability density function (PDF)~\cite{crystalball}.
In order to determine the shape parameters of this PDF, we first fit a larger sample selected without
tagging of the one-prong side. Using this fixed shape, we then estimate the number of $\fivepipizero$
events within our signal region ($1.3<M^{*}<1.8\,\gevcc$) from the MC sample with the one-prong tag applied.
We obtain 3.6$\pm$0.6 events, scaled to the luminosity of 232\,$\invfb$, where the
uncertainty is statistical only (see Figure~\ref{fig:fitbkg}, left).
Simply counting the number of events in the signal region yields 3.2 (scaled) MC events.

The uncertainty of the $\fivepipizero$ background estimate is based on the uncertainties of the fitted 
PDF shape parameters,
namely, the central value and the width, and the correlation between them. 
The values of the PDF shape parameters are randomly generated 
according to their uncertainties expressed in the covariance matrix,
and the resulting PDF is then used to estimate the number of background events in the signal region.
The total uncertainty from the fitting (0.6 events, 16.7\,\%) is added in quadrature with systematic
uncertainties in the reconstruction of the tracks and neutrals, particle identification, luminosity
and $\tau$-pair cross section (8.4\,\%) and the uncertainty in the branching ratio
of the $\fivepipizero$ decay mode (14.9\,\%).

An additional background contribution is expected from the $\twopizero$ mode.
Using an MC sample corresponding to 675\,$\invfb$ of data we estimate 0.7$\pm$0.5 
background events in the signal region from this source.
The uncertainty is dominated by the MC statistics.
Contributions from other generic $\tau$ decays are negligible.
Combining both sources of the $\tau$ background, we expect
a total of $4.3\pm1.0$ background events in the data.

For this analysis, a comparison of MC simulation and data has shown that
the $\eetoqq$ background contributions cannot reliably be extracted from
simulation due to difficulties in modeling the fragmentation processes. 
The shape of the simulated pseudo-mass distribution appears to agree
with the shape in the data, but the overall normalization does not.
Therefore, the $q\ov{q}$ background is estimated directly from the data,
by fitting the data pseudo-mass spectrum with the sum of two
Gaussians. This PDF is motivated by MC studies, which show that the
$\eetouds$ and $\eetocc$ backgrounds have Gaussian pseudo-mass shapes with different
parameters. The double-Gaussian fit to the MC pseudo-mass distribution of $q\ov{q}$ background 
is shown in Figure~\ref{fig:fitbkg} (right). 

To extract the $q\ov{q}$ background in the signal region, we subtract the
expected $\tau$ background contribution from the data pseudo-mass
distribution, and fit the resulting histogram in the range
$1.8<M^*<3.3\,\gevcc$ with a double-Gaussian PDF whose means and sigmas are allowed
to float. To avoid experimenter bias, this fit is performed ``blind'', with the
data in the signal region hidden. The fit function is then extrapolated
below $1.8\,\gevcc$ and its integral between $1.3$ and 
$1.8\,\gevcc$ yields the $q\ov{q}$ background estimate in the data,
$2.2$ events.

To calculate the statistical uncertainty of the $q\ov{q}$ background estimate we vary the number of events in each bin of 
the data $q\ov{q}$ pseudo-mass spectrum above $1.8\,\gevcc$ according to its Poisson error and 
refit the resulting histogram for a new estimate. The statistical uncertainty of $^{+1.6}_{-1.0}$ events is 
extracted from the 
variance of the distribution of the generated $q\ov{q}$ background estimates.
Variations in the functional form of the fit PDF are taken into account as a systematic uncertainty of $^{+0.7}_{-0.0}$ 
events. The total uncertainty is calculated by adding the statistical and systematic uncertainties in quadrature. 
Thus, the $q\ov{q}$ background estimate is $2.2^{+1.7}_{-1.0}$ events.  

To validate the $\eetoqq$ background estimate method, we apply it to a
$\tau$-event-free data sample, obtained by requiring at least 3 photons with energies greater than $300\,\mev$ on the
tag side not associated with a $\pi^0$. 
This requirement effectively suppresses $\tau$ events to a negligible
level and provides a clean $q\ov{q}$ sample in the data. Comparison between
the expected and observed $q\ov{q}$ background levels for this sample shows good agreement, 
11.8 predicted background events vs. 12 observed.

Another cross-check we perform is the branching ratio measurement of the $\fivepipizero$ decay mode
using the same selection criteria (except for demanding only one
$\pi^0$ on the signal side instead of two) as described above. The measured branching ratio is consistent
with the Particle Data Group's value~\cite{pdg04}. 

\begin{figure}[thb]
\begin{center}
\epsfig{file=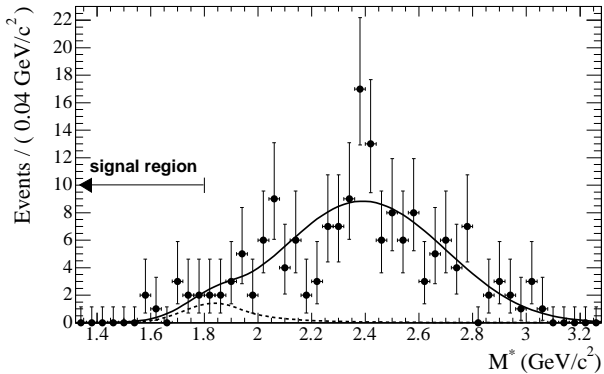,width=0.47\textwidth}
\caption{\small{Pseudo-mass distribution of the data events passing the $\signal$ selection criteria.
The solid curve represents the total expected background PDF. 
The dashed curve illustrates the $\tau$ background contribution.
\label{fig:results}}} 
\end{center}
\end{figure}

Combining the background estimates from $\tau$ and $q\ov{q}$ events,
we calculate a total of $6.5^{+2.0}_{-1.4}$ background events.
Figure~\ref{fig:results} illustrates the final pseudo-mass spectrum of the data, along with
the expected background PDF. We observe $10$ events in the signal region
and conclude that there is no evidence for the $\signal$ decay.

The upper limit for the $\signal$ decay branching ratio is calculated from
\begin{equation}
\BR(\signal) < \frac{\lambda_{N_{\rm signal}}}{2 \times N_{\tau\tau} \times \epsilon \label{eq:limit}}~,
\end{equation}
where $\lambda_{N_{\rm signal}}$ is the upper limit on the number of signal events at the 90\,\% CL.
This number is obtained using a limit calculator program~\cite{BARLOW} that follows the Cousins and
Highland approach~\cite{COUSINS} of incorporating systematic uncertainties into the upper limit, 
using the numbers of expected background and observed events, as well as the uncertainties on the
background, signal efficiency and the number of $\tau$ pairs. We find 
$\lambda_{N_{\rm signal}}=9.2$ events and \BR$(\signal) < 3.4\times10^{-6}$ at the 90\,\% CL. 
Table~\ref{tab:results} summarizes the results of this analysis.
\begin{table}[hbt]
\caption{Signal efficiency, expected background, observed data events, and the upper limit of the 
$\signal$ decay at the 90\,\% CL. \label{tab:results}}
\renewcommand{\arraystretch}{1.2}
\begin{tabular}{lc}
\hline
\hline
\normalsize $N_{\tau\tau}$	 & \normalsize (206.5$\pm$4.7)$\times$10$^6$ \\
\hline
\normalsize $\signal$ efficiency   & \normalsize (0.66$\pm$0.05)\,\%  \\
\normalsize Expected $\tau^{+}\tau^{-}$ background  & \normalsize 4.3$\pm$1.0 events \\
\normalsize Expected $q\ov{q}$  background		& \normalsize 2.2$^{+1.7}_{-1.0}$ events\\
\normalsize Expected total 	background		& \normalsize 6.5$^{+2.0}_{-1.4}$ events \\
\normalsize Observed events				& \normalsize 10 \\
\hline
\normalsize \BR$(\signal)$ &\normalsize  $<$ 3.4 $\times$ 10$^{-6}$ \\
\hline
\hline
\end{tabular}
\end{table}

In addition to this inclusive result, we also search for the resonant
decay mode $\twoomegapi$ with the subsequent decay $\omegadec$, which is  
predicted to be the main channel for the
$\signal$ decay~\cite{Sobie}. The $\twoomegapi$ mode has a much
narrower allowed pseudo-mass range ($1.7<M^{*}<1.8\,\gevcc$) due to its kinematics.
For the same reason, the background level is expected to be much smaller.
The event selection is re-optimized for this analysis.
Photons are required to have a minimum energy 
of $50\,\mev$, energy deposited in at least two crystals and a lateral energy profile consistent with that of a photon.
Reconstructed $\pi^0$ candidates must have energies above $200\,\mev$. 
The $\omega$ resonance is reconstructed as a $\pi^+\pi^-\pi^0$ combination with an invariant 
mass of $0.76<M_{\pi^+\pi^-\pi^0}<0.80\,\gevcc$. 

Reconstruction of both $\omega$ mesons suppresses the background and therefore further selection cuts can be 
substantially loosened to increase the signal efficiency.
The conversion veto and the $E_{\rm res}$ cuts are not used.
In addition, we allow one charged particle of any type on the tag side, 
and only loose pion identification
is required on the signal side. As a result, the $\twoomegapi$ efficiency for this selection
is (1.53$\pm$0.13)\,\%. The uncertainty is a combination of systematic and statistical 
uncertainties, as described above for the inclusive $\signal$ analysis.

\begin{figure}[thb]
\begin{center}
\epsfig{file=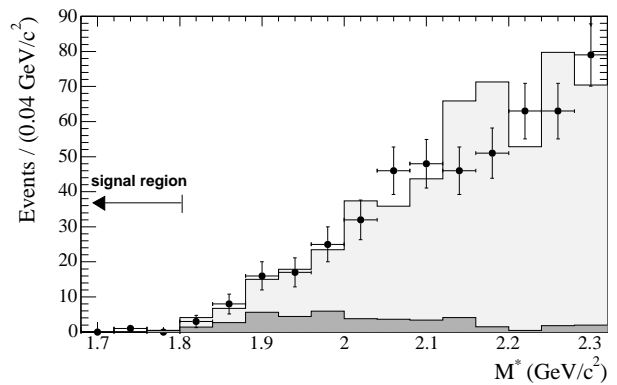,width=0.47\textwidth}
\caption{\small{Pseudo-mass distributions of the data (points) and MC (shaded histograms) events
passing the $\twoomegapi$ selection criteria. The dark shaded histogram corresponds to the $\tau$ background, whose
level is determined from the simulation. The light histogram
shows the total background, with the level of the $q\ov{q}$ contribution
scaled to agree with the data.
The data signal region below $1.8\,\gevcc$ was blinded during the background estimation. \label{fig:2omegapi}}} 
\end{center}
\end{figure}

The background is estimated from MC simulation (see Figure~\ref{fig:2omegapi}).
As in the inclusive analysis, while there is a discrepancy between
the data and MC $q\ov{q}$ yields, the shape of the MC $q\ov{q}$ pseudo-mass spectrum agrees
with the data.
As a result of the study we expect negligible $q\ov{q}$ contribution in the
signal region.
The uncertainty on the $q\ov{q}$ background estimate is calculated using the same technique described
for the inclusive $\signal$ analysis. The total expected $q\ov{q}$ background is 0.0$^{+0.1}_{-0.0}$ events.
An additional contribution comes from the $\omegathreepi$ mode. Out of 530\,$\invfb$ of MC simulated
$\omegathreepi$ events, only 1 event is found in the signal region. Thus, we expect 0.4$^{+1.0}_{-0.4}$
events in 232\,$\invfb$ of data. The uncertainty in the $\tau$ background estimate is calculated as a Poisson error of 1 event at 68\,\% CL.

We find 1 event passing the selection criteria in 232\,$\invfb$ of data, which is consistent with the expected background.
We calculate the upper limit of the $\twoomegapi$ decay branching ratio
using the limit calculator~\cite{BARLOW}, which yields $\lambda_{N_{\rm signal}}=3.4$ events
at the 90\,\% CL. 
The upper limit for the decay, \BR$(\signal) < 5.4\times10^{-7}$, is significantly lower than for the inclusive 
decay $\signal$. Table~\ref{tab:2omegapi_results} summarizes the results of the $\twoomegapi$ search.
\begin{table}[hbt]
\caption{Signal efficiency, expected background, observed data events, and the upper limit of the 
$\twoomegapi$ decay at the 90\,\% CL. \label{tab:2omegapi_results}}
\renewcommand{\arraystretch}{1.2}
\begin{tabular}{lc}
\hline
\hline
\normalsize $N_{\tau\tau}$	 & \normalsize (206.5$\pm$4.7)$\times$10$^6$ \\
\hline
\normalsize $\twoomegapi$ efficiency   & \normalsize (1.53$\pm$0.13)\,\%  \\
\normalsize Expected $\tau^{+}\tau^{-}$ background  & \normalsize 0.4$^{+1.0}_{-0.4}$ events \\
\normalsize Expected $q\ov{q}$  background		& \normalsize 0.0$^{+0.1}_{-0.0}$ events \\
\normalsize Expected total 	background		& \normalsize 0.4$^{+1.0}_{-0.4}$ events\\
\normalsize Observed events				& \normalsize 1 \\
\hline
\normalsize \BR$(\twoomegapi)$ & \normalsize $<$ 5.4 $\times$ 10$^{-7}$ \\
\hline
\hline
\end{tabular}
\end{table}

In conclusion, we present results of a search for the $\signal$ and
$\twoomegapi$ decay modes using 232\,$\invfb$ of data collected
by the \babar\ detector. No evidence for these decays is found.
We calculate \BR$(\signal) < 3.4\times10^{-6}$ at the 90\,\% CL,
improving the existing experimental limit for this mode by more than a factor of 30.
The upper limit for the decay, \BR$(\twoomegapi) < 5.4\times10^{-7}$,
is reported here for the first time. \\

\input pubboard/acknowledgements.tex

\end{document}

%% file: pubboard/authors_feb2006.tex
%
\author{B.~Aubert}
\author{R.~Barate}
\author{M.~Bona}
\author{D.~Boutigny}
\author{F.~Couderc}
\author{Y.~Karyotakis}
\author{J.~P.~Lees}
\author{V.~Poireau}
\author{V.~Tisserand}
\author{A.~Zghiche}
\affiliation{Laboratoire de Physique des Particules, F-74941 Annecy-le-Vieux, France }
% ...
\collaboration{The \babar\ Collaboration}
\noaffiliation

%% file: pubboard/acknowledgements.tex
We are grateful for the 
extraordinary contributions of our \pep2\ colleagues in
achieving the excellent luminosity and machine conditions
that have made this work possible.
The success of this project also relies critically on the 
expertise and dedication of the computing organizations that 
support \babar.
The collaborating institutions wish to thank 
SLAC for its support and the kind hospitality extended to them. 
This work is supported by the
US Department of Energy
and National Science Foundation, the
Natural Sciences and Engineering Research Council (Canada),
Institute of High Energy Physics (China), the
Commissariat \`a l'Energie Atomique and
Institut National de Physique Nucl\'eaire et de Physique des Particules
(France), the
Bundesministerium f\"ur Bildung und Forschung and
Deutsche Forschungsgemeinschaft
(Germany), the
Istituto Nazionale di Fisica Nucleare (Italy),
the Foundation for Fundamental Research on Matter (The Netherlands),
the Research Council of Norway, the
Ministry of Science and Technology of the Russian Federation, and the
Particle Physics and Astronomy Research Council (United Kingdom). 
Individuals have received support from 
CONACyT (Mexico), the Marie-Curie Intra European Fellowship program (European Union),
the A. P. Sloan Foundation, 
the Research Corporation,
and the Alexander von Humboldt Foundation.

%% file: paper.bbl
\begin{thebibliography}{3}
\bibitem{footnote} Throughout this paper, whenever a mode is given its charge conjugate is also implied.
\end{thebibliography}
